\documentclass{ifacconf}
\usepackage{graphicx}      
\usepackage{natbib}
\usepackage{caption}
\usepackage{amsmath} 
\usepackage{amssymb}  

\usepackage{color}
\usepackage{graphics} 
\usepackage{epsfig} 
\usepackage{subfigure}
\usepackage{ulem}
\usepackage{tikz,pgfplots}
\usetikzlibrary{patterns}
\usepackage{algorithmic}
\usepackage{algorithm}

\newcommand{\lefri}[1]{\left(#1\right)}

\newcommand{\dt}[0]{\mbox{d}t}

\newcommand{\CQ}[0]{C^{2}\lefri{\left[0,h\right],Q}}

\newcommand{\truqargsf}[2]{\substack{q \in \CQ\\q\lefri{0} = #1, q\lefri{h} = #2}}

\newcommand{\ext}[0]{\operatornamewithlimits{extremize}}

\usepackage{amssymb}

\newcommand{\proa}{A^*G \mbox{$\;$}_{\tau^*} \kern-3pt\times_\alpha
G \mbox{$\;$}_\beta \kern-3pt\times_{\tau^*} A^*G}

\tikzstyle{vertex}=[circle,fill=black!20,minimum size=15pt,inner sep=0pt]
\tikzstyle{selected vertex} = [vertex, fill=red!24]
\tikzstyle{edge} = [draw,thick,-]
\tikzstyle{dedge} = [draw,thick,<->]
\tikzstyle{shadowdedge} = [draw, dotted,->]
\tikzstyle{weight} = [font=\small]
\tikzstyle{selected edge} = [draw,line width=3pt,-,red!50]
\tikzstyle{ignored edge} = [draw,line width=3pt,-,black!20]

\begin{document}

\begin{frontmatter}

\title{Forced variational integrator for distance-based shape control with flocking behavior of multi-agent systems\thanksref{footnoteinfo}} 

\thanks[footnoteinfo]{L. Colombo, H. Garc\'ia de Marina and M. Cao were partially supported by I-Link Project (Ref: linkA20079) from CSIC. L. Colombo was partially supported by Ministerio de Economia,
Industria y Competitividad (MINEICO, Spain) under grant MTM2016-
76702-P; ''Severo Ochoa Programme for Centres of Excellence'' in R$\&$D
(SEV-2015-0554). The project that gave rise to these results received the
support of a fellowship from ``la Caixa'' Foundation (ID 100010434). The
fellowship code is LCF/BQ/PI19/11690016. P. Moreno is supported by a Peruilh fellowship from the School of Engineering from the University of Buenos Aires. The work of Hector Garcia de Marina is supported by the grant Atracci\'on de Talento 2019-T2/TIC-13503 from the Government of the Autonomous Community of Madrid. M. Cao is supported by the European Research Council (ERC-CoG-771687).}

\author[First]{Leonardo Colombo}
\author[Second]{Patricio Moreno}
\author[Third]{Mengbin Ye}
\author[Forth]{H\'ector Garcia de Marina}
\author[Fifth]{Ming Cao}

\address[First]{Institute of Mathematical Sciences (ICMAT), Madrid, Spain (e-mail: leo.colombo@icmat.es).}
\address[Second]{LAR-GPSIC, Facultad de Ingenier\'ia, Universidad de Buenos Aires, Argentina (e-mail: pamoreno@fi.uba.ar)}
\address[Third]{Optus-Curtin Centre of Excellence in Artificial Intelligence, Curtin University, Australia (mengbin.ye@curtin.edu.au).}
\address[Forth]{ Faculty of Physics, Department of Computer Architecture and Automatic Control,
Universidad Complutense de Madrid, Madrid, Spain (hgdemarina@gmail.com.).}
\address[Fifth]{Faculty of Science and Engineering,
ENTEG, University of Groningen, Netherlands (m.cao@rug.nl).}

\begin{abstract}
A multi-agent system designed to achieve distance-based shape control with flocking behavior can be seen as a mechanical system described by a Lagrangian function and subject to additional external forces. Forced variational integrators are given by the discretization of Lagrange-d'Alembert principle for systems subject to external forces, and have proved useful for numerical simulation studies of complex dynamical systems. We derive forced variational integrators that can be employed in the context of control algorithms for distance-based shape with velocity consensus. In particular, we provide an accurate numerical integrator with a lower computational cost than traditional solutions, while preserving the configuration space and symmetries. We also provide an explicit expression for the integration scheme in the case of an arbitrary number of agents with double integrator dynamics. For a numerical comparison of the performances, we use a planar formation consisting of three autonomous agents.
  \end{abstract}


\end{frontmatter}

\section{Introduction}

In many engineering applications, numerical integrators for continuous-times equations of motion of physical systems are usually derived by discretizing differential equations. However, the inherent geometric structure of the governing continuous-time equations and conserved quantities are not preserved in simulations with the traditional integrators. Variational integrators are numerical methods derived from the discretization of variational principles as it has been surveyed by \cite{Hair, mawest}. These integrators retain some of the key geometric properties of the continuous systems, such as symplecticity, momentum conservation, and also exhibit easily verifiable behavior of the energy associated to the system. This class of numerical methods have been applied to a wide range of problems in optimal control, constrained systems, power systems, nonholonomic systems, and systems on Lie groups \cite{ober2011discrete,leyendecker2010discrete,cortes2002geometric,kobilarov2011discrete}.

The past two decades have seen a great advance in the development of algorithms for the coordination of multi-agent systems \cite{anderson, Oh}. The development of new integration schemes to implement these algorithms has reliable crucially on accurate and fast simulations to numerically determine regions of attraction in swarms, as well as, enable more computationally efficient estimation algorithms like Kalman filters that employ distance-based controllers as prediction models. We have observed in \cite{CoGa} that variational integrators help actual implementations of distributed multi-agent systems in formation control by relaxing the requirements in computational cost (energy
efficiency) per agent as much as possible. In particular,
agents can employ variational integrators for their estimation algorithms to save energy consumption, having a lower computational cost than traditional numerical solutions like Runge-Kutta, and
without compromising accuracy (Euler integrator). Moreover, while the Runge-Kutta scheme is a multi-step method, for multi-agent systems with a double integrator dynamics, variational integrators can be one-step algorithms. 

In our previous work \cite{CoGa}, we showed how variational integrators can be used for formation shape control with null final velocity. That is, given a set of initial conditions the agents move along the work-space and achieve a prescribed formation shape with final velocity equal to zero, meaning they do not follow a motion keeping the formation. In contrast, in this work we exploit the properties of controllers for distance-based shape control with velocity consensus for agents described by double integrator dynamics as in \cite{fl1,Di,Oh} to construct forced variational integrators for this unstudied situation in \cite{CoGa}. The goal of this work is to derive forced variational integrators that can be employed in the context of distance-based shape control algorithms with velocity consensus presenting more accurate qualitative features compared to traditional integrators. As a result, we employ the variational integrators for high accuracy numerical solutions without compromising the computational cost. In fact, multi-agent systems can consist of a significant number of agents and links (i.e. neighboring agents) where the larger the set of initial conditions, the greater the sensitivity for the agents' trajectories. The employment of the proposed integrator shows clear advantages exhibiting the accuracy of a Runge-Kutta method yet with the low computational cost of an explicit Euler method. Moreover, the behavior of transitory shapes generated by the variational integrator improves the ones provided for instance, by an explicit Euler method. The integrator presented in this work also preserves the configuration space, symmetries, shows a good behavior of the energy dissipated along the motion, and it provides an accurate numerical scheme with a lower computational cost than traditional solutions.

In this paper, we introduce a mathematical framework to study formation control of multiple Lagrangian systems and we construct a geometric integrator based on the discretization of an extension of the Lagrange-d'Alembert principle for a single agent, in the spirit of forced variational integrators \cite{mawest}. This is because distance-based shape control with flocking behavior of multiple mechanical systems can be seen as a physical system of particles linked by springs, whose evolution can be described by a Lagrangian function subject to conservative forces coming from the potential whose minimum corresponds to the desired distance-based shape, and external dissipative forces coming from the velocity consensus between the agents. The new situation studied in this work, compared with \cite{CoGa} needs the development of a new variational principle and the consideration of external forces where more than one agent is involved.

The structure of the paper is as follows. Section $2$ introduces Lagrangian mechanics and variational integrators. In Section $3$, we describe shape control with flocking behavior for multi-agent systems as a Lagrangian system subject to external forces and we derive by a variational principle the corresponding equations of motion. In Section $4$ we discretize the variational principle given in Section $3$ and we derive a forced-variational integrator for distance-based shape control with velocity consensus. Section $5$ gives a numerical comparison and a discussion of the numerical method developed in this work against classical numerical integrators.

\section{Variational Integrators}
\subsection{Lagrangian mechanics}

Let $Q$ be an $n$-dimensional differentiable manifold, the
configuration space of a mechanical system, and denote by $(q^A)$, $1\leq A\leq n$, local coordinates on $Q$. Denote by $TQ$ its
tangent bundle, that is $\displaystyle{TQ=\bigcup_{q\in Q}T_{q}Q}$ with induced local coordinates $(q^A, \dot{q}^A)$.  $T_{q}Q$ denotes the tangent space of $Q$ at the point $q$. $T_qQ$ has a vector space structure, so we may consider its dual space, $T^{*}_{q}Q$. The cotangent bundle $T^{*}Q$ is defined as $\displaystyle{T^{*}Q=\bigcup_{q\in Q}T^{*}_{q}Q}$, with induced local coordinates $(q^A,p_A)$.

Given a Lagrangian function $L:TQ\rightarrow \mathbb{R}$, its Euler-Lagrange
equations are
\begin{equation}\label{qwer}
\frac{d}{dt}\left(\frac{\partial L}{\partial\dot
q^A}\right)-\frac{\partial L}{\partial q^A}=0, \quad 1\leq A\leq n.
\end{equation}
These equations determine a system of implicit second-order
differential equations. If we assume that the Lagrangian is \textit{regular},
that is, the ${n\times n}$ matrix $\left(\frac{\partial^{2} L}{\partial \dot q^A
\partial \dot q^B}\right)$, with $A, B=1,\ldots,n$, is non-singular, then the local existence and uniqueness of solutions is guaranteed for any given initial condition.

\subsection{Variational Integrators}\label{sec: div} 

A \textit{discrete Lagrangian} is a differentiable function
$L_d\colon Q \times Q\to\mathbb{R}$, which may be considered as an
approximation of the action integral defined by a continuous regular 
Lagrangian $L\colon TQ\to \mathbb{R}$ over the time step $\left[0,h\right]$. Given a  small time step $h>0$,
\[
L_d(q_0, q_1, h)\approx \int^h_0 L(q(t), \dot{q}(t))\; dt,
\]
where $q(t)$ is the unique solution of equation \eqref{qwer} with  boundary conditions $q(0)=q_0$ and $q(h)=q_1$. That is, \begin{align*}
L_{d}\lefri{q_{0},q_{1},h} \approx \ext_{\truqargsf{q_{0}}{q_{1}}} \int_{0}^{h} L\lefri{q,\dot{q}} \dt.
\end{align*}%

From now on we will write \(L_{d}\lefri{q_{0},q_{1}}\) when \(h\) is assumed to be constant. We construct the grid $\mathcal{T}=\{t_{k}=kh\mid k=0,\ldots,N\},$ with $Nh=T$, with $T$ being the total time of interest in developing the integrator,
and also define the discrete path space
$\mathcal{P}_{d}(Q):=\{q_{d}:\{t_{k}\}_{k=0}^{N}\to Q\}.$ We
identify a discrete trajectory $q_{d}\in\mathcal{P}_{d}(Q)$ with its
image $q_{d}=\{q_{k}\}_{k=0}^{N}$, where $q_{k}:=q_{d}(t_{k})$. The
discrete action $\mathcal{A}_{d}:\mathcal{P}_{d}(Q)\to\mathbb{R}$ for this
sequence of discrete paths is calculated by summing the discrete Lagrangian on each
adjacent pair, and it is defined by
\begin{equation}\label{acciondiscreta}
\mathcal{A}_d(q_{d}) = \mathcal{A}_d(\{q_k\}_{k=0}^{N})=\sum_{k=0}^{N-1}L_d(q_k,q_{k+1}).
\end{equation}
Note that the discrete path space is
isomorphic to the smooth product manifold which consists of $(N+1)$
copies of $Q$. The discrete action inherits the smoothness of the
discrete Lagrangian and the tangent space
$T_{q_{d}}\mathcal{P}_{d}(Q)$ at $q_{d}$ is the set of maps
$v_{q_{d}}:\{t_{k}\}_{k=0}^{N}\to TQ$ such that $\tau_{Q}\circ
v_{q_{d}}=q_{d}$. 


The discrete variational principle states that
the solutions of the discrete system determined by $L_d$ must
extremize the action sum given fixed points $q_0$ and $q_N.$
Minimizing $\mathcal{A}_d$ over $q_k$ with $1\leq k\leq N-1,$ we
obtain the following system of difference equations
\begin{equation}\label{discreteeq}
 D_1L_d( q_k, q_{k+1})+D_2L_d( q_{k-1}, q_{k})=0,
\end{equation} where $D_j$ stands for the partial derivative with respect to the $j$-th component of $L_d$. These equations are called \textit{discrete Euler-Lagrange
equations}, and the reader may compare this to \eqref{qwer}. 

Given a solution $\{q_{k}^{*}\}_{k\in\mathbb{N}}$ of
Eq.\eqref{discreteeq} and assuming that the
matrix $(D_{12}L_d(q_k, q_{k+1}))$ is non-singular (regularity hypothesis), it is possible to
define implicitly a (local) discrete flow $
\Upsilon_{L_d}\colon\mathcal{U}_{k}\subset Q\times Q\to Q\times Q$
by $\Upsilon_{L_d}(q_{k-1}, q_k)=(q_k, q_{k+1})$ from
(\ref{discreteeq}), where $\mathcal{U}_{k}$ is a neighborhood of the
point $(q_{k-1}^{*},q_{k}^{*})$.

The exact solution $q(t)$ for the boundary value problem in equation \eqref{qwer} is in general not known in exact form, so one must consider approximations of the trajectory $q(t)$. The general idea for the construction of $L_d$ is as follow. Let $L\colon TQ \to \mathbb{R}$ and $[0,T]$ be given. Divide $[0,T]$ into $N$ pieces of size $h=T/N$ (time step). If $Q$ is for instance a vector space, consider the approximation 
$\displaystyle{
q(t)\approx \frac{q_0+q_1}{2}}$ and $\displaystyle{\dot q(t)\approx \frac{q_1 - q_0}{h}}$,
 which enables us to define
\begin{align*}
L_d(q_0,q_1)=&\int_0^hL\left(\frac{q_0+q_1}{2}, \frac{q_1 - q_0}{h}\right)dt\\
=&hL\left(\frac{q_0+q_1}{2}, \frac{q_1 - q_0}{h}\right).
\end{align*}



\section{Distance-based shape control with flocking behavior for Lagrangian systems}


Consider a set $\mathcal{V}$ consisting of $s\geq 2$ free agents evolving each one on $Q$. We denote by $q_i\in Q$ the configurations (positions) of agent $i\in\mathcal{V}$, with local coordinates $q_i^{A}=(q_i^{1},\ldots,q_i^{n})$, and by $q=(q_1,\ldots,q_s)\in Q^{s}$ the stacked vector of positions where $Q^{s}$ represents the Cartesian product of $s$ copies of $Q$.

For simplicity in the exposition, we study the case where the neighbor relationships between agents are described by an undirected and unweighted graph $\mathcal{G}=(\mathcal{V},\mathcal{E})$, without self loops, where $\mathcal{V}$ denotes the set of nodes and the set $\mathcal{E}\subset\mathcal{V}\times\mathcal{V}$ denotes the set of un-ordered edges of the graph. We assume that the graph is static and connected. 

The set of neighbors for agent $i$ is defined by $\mathcal{N}_i=\{j\in\mathcal{V}: (i,j)\in\mathcal{E}\}$. For shape control we define the incidence matrix $B\in Q^{s\times|\mathcal{E}|}$ for $\mathcal{G}$ by 
$\displaystyle{
b_{ik}=
\begin{cases}
+1 \hbox{ if } i=\mathcal{E}_k^{tail},\\
-1 \hbox{ if } i=\mathcal{E}_k^{head}\\
0 \hbox{ otherwise }
\end{cases}
\label{eq: B}
}$ where $\mathcal{E}_{k}^{tail}$ and $\mathcal{E}_k^{head}$ denote the tail and head nodes, respectively, of the edge $\mathcal{E}_k$, i.e., $\mathcal{E}_k=(\mathcal{E}_k^{tail},\mathcal{E}_k^{head})$.

In this work, the motion of each agent will be described as a Lagrangian system on $TQ$, that is, the motion of the agent $i\in\mathcal{V}$ is described by the Lagrangian function $L_i:TQ\to\mathbb{R}$ and the dynamical system associated with $L_i$ is given by the Euler-Lagrange equations, i.e., $$\frac{d}{dt}\left(\frac{\partial L_i}{\partial\dot{q}_i^A}\right)-\frac{\partial L_i}{\partial q_i^A}=0,\,i\in\mathcal{V} \hbox{ and } A=1,\ldots,n.$$


In addition, the agent $i\in\mathcal{V}$  may be influenced by a non-conservative force (conservative forces may be included into the potential energy $V_i$), which is a smooth map $F_{ij}:TQ\times TQ\to T^{*}Q$. For instance, $F_{ij}$ can describe a velocity consensus algorithm, that is, how each agent adjust its velocity with respect of its neighbor $j\in\mathcal{N}_i$. 

At a given position and velocity, the force will act against variations of the position (virtual displacements). A consequence  of  the Lagrange-d'Alembert principle or principle of virtual work (see \cite{Bloch}) establishes that the natural motions of the system are those paths $q:[0,T]\to Q$ satisfying 
\begin{equation}\label{dlp}\delta\int_{0}^{T}L_i(q_i,\dot{q}_i)\,dt+\int_{0}^{T}F_{ij}(q_i,q_j,\dot{q}_i,\dot{q}_{j})\delta q_i\,dt=0\end{equation} for all variations satisfying $\delta q_i(0)=\delta q_i(T)=0$. The second term in \eqref{dlp} is known as virtual work since $F_{ij}(q_i,q_j,\dot{q}_i,\dot{q}_j)\delta q_i$ is the virtual work done by the force field $F_{ij}$ with a virtual displacement $\delta q_i$. The Lagrange-d'Alembert principle leads to the forced Euler-Lagrange equations \begin{equation}\label{feleq}\frac{d}{dt}\left(\frac{\partial L_i}{\partial\dot{q}_i^A}\right)-\frac{\partial L_i}{\partial q_i^A}=F_{ij}(q_i,q_j,\dot{q}_i,\dot{q}_j).\end{equation} 


Consider the Lagrangian function $\mathbf{L}:(TQ)^{s}\to\mathbb{R}$ for the multi-agent system defined by \begin{equation}\label{lagrangianL}\mathbf{L}(q,\dot{q})=\sum_{i=1}^{s}L_i(\pi_i(q),\tau_i(\dot{q}))\end{equation} where $L_i:TQ\to\mathbb{R}$ is the Lagrangian for the agent $i\in\mathcal{V}$, $\displaystyle{(TQ)^{s}=\Pi_{i=1}^{s}TQ}$, $\pi_{i}:Q^{s}\to Q$ is the canonical projection from $Q^{s}$ over its $i^{th}$-factor and $\tau_i:(TQ)^{s}\to TQ$ is the canonical projection from $(TQ)^{s}$ over its $i^{th}$-factor. That is, $\pi_i(q)=q_i\in Q$ and $\tau_i(q,\dot{q})=(q_i,\dot{q}_i)$ with $(q,\dot{q})\in (TQ)^{s}$.
 
 
To control the shape of the formation we introduce the artificial potential functions $V_{ij}:Q\times Q\to\mathbb{R}$ for $i,j\in\mathcal{V}$ and $i\neq j$
\begin{equation}
	V_{ij}(q_i,q_j)=\frac{1}{4}(||q_{ij}||^2-d_{ij}^2)^2,
	\label{eq: Vij}
\end{equation}
where $||\cdot||$ is a norm on $Q$ inducing a distance, $q_{ij}$ denotes the relative position between agents $i$ and $j$, and $d_{ij}$ is the desired distance between agents $i$ and $j$. If we are interested in stabilizing a specific geometrical shape, then the incidence matrix $B$ and the set of desired distances can be determined by the rigidity theory as reviewed by \cite{anderson}, \cite{fl1}. 

By flocking behavior we mean that all agents achieve a consensus in the velocities. Flocking behavior can be achieved by means of the Laplacian matrix associated with $\mathcal{G}$. The Laplacian matrix $\mathcal{L}$ is the matrix whose entries are given by $l_{ij}=-1$ with $i\neq j$, if there is an edge between agents $j$ and $i$, else $l_{ij}=0$. Moreover, $\displaystyle{l_{ii}=-\sum_{j\in\mathcal{N}_{i}}l_{ij}}$.  In the case of $\mathcal{G}$ being an undirected graph, it follows that $\mathcal{L}=BB^{T}$.

Define $\overline{\mathcal{L}}=\mathcal{L}\otimes I_n$, then the consensus algorithm describing how each agent adjusts its velocity is given by \begin{equation}\label{fll}\ddot{q}=-\overline{\mathcal{L}}\dot{q}.\end{equation} That is, for agent $i$, equation \eqref{fll} is equivalent to $$\ddot{q}_{i}=-\sum_{j\in\mathcal{N}_i}l_{ij}(\dot{q}_i-\dot{q}_j).$$ These equations corresponds with  the forced Euler Lagrange equations \eqref{feleq} with $L_i=\frac{1}{2}||\dot{q}_i||^2$ and force given by $F_{ij}=-\sum_{j\in\mathcal{N}_i}l_{ij}(\dot{q}_i-\dot{q}_j)$. Under these conditions the Lagrangian for the formation problem $\mathbf{L}_F:(TQ)^s\to\mathbb{R}$ takes the form \begin{equation}\label{lagrangianLV}\mathbf{L}_{F}(q,\dot{q})=\sum_{i=1}^{s}(\underbrace{L_i(\pi_i(q),\tau_i(\dot{q}))+\frac{1}{2}\sum_{j\in\mathcal{N}_i}V_{ij}(\pi_i(q),\pi_j(q))}_{\mathcal{L}_{i}(q,\dot{q})}).\end{equation}
 

  \begin{prop}\label{Th1}
Let $\mathbf{L}_{F}:(TQ)^s\to\mathbb{R}$ be the Lagrangian function defined in \eqref{lagrangianLV} and $F:(TQ)^s\to(T^{*}Q)^s$ be external forces. The curve
$q\in\mathcal{C}^{\infty}(Q^s)$ satisfies
$\delta\mathcal{A}(q)=0$ for the action functional defined by \begin{equation*}\mathcal{A}(q)\!=\!\!\int_{0}^{T}\!\!\sum_{i=1}^{s}(\mathcal{L}_{i}(q,\dot{q})+\!\sum_{j\in\mathcal{N}_i}\!\!F_{ij}(\pi_i(q),\pi_j(q),\tau_i(\dot{q}), \tau_j(\dot{q})))dt\end{equation*}
if and only if, for variations of $q\in Q^s$ with fixed endpoints and the virtual work done by the forces when the path $q(t)$ is only varied by $\delta q(t)$, $q$ is a solution of the
forced Euler-Lagrange equations for $\mathbf{L}_F$:\begin{equation}\label{eqforces}
\frac{d}{dt}\left(\frac{\partial L_i}{\partial \dot{q}_i^A}\right)-\frac{\partial L_i}{\partial q_i^A}=\sum_{j\in\mathcal{N}_i} \left(F_{ij}(q_i,q_j,\dot{q}_i,\dot{q}_j)-\frac{\partial V_{ij}}{\partial q_i^A}\right), \end{equation}for all $A=1,\ldots,n \hbox{ and  for each } i\in\mathcal{V}$.
\end{prop}

\section{Variational Integrator for distance-based shape control with flocking behavior}

The key idea of variational integrators is that discretization occurs for the variational principle rather than the resulting equations of motion. As  in Section \ref{sec: div}, we discretize the state space $TQ$ as $Q\times Q$. For each agent $i\in\mathcal{V}$, consider a discrete Lagrangian $L_{i}^d:Q\times Q\to\mathbb{R}$ and discrete ``external forces" $F_{ij,d}^{\pm}:(Q\times Q)\times (Q\times Q)\to T^{*}Q$ approximating the integral action and the virtual work, respectively,  as 
 \begin{align}
 \int_{t_k}^{t_{k+1}}L_i(q_i,\dot{q}_i)\,dt\simeq &L_{i}^d(q_{k}^{i},q_{k+1}^{i}),\label{eqq1}\\
\int_{t_k}^{t_{k+1}}F_{ij}(q_i,q_j,\dot{q}_i,\dot{q}_j)\delta q_i\,dt&\simeq F_{ij,d}^{-}(q_{k}^{i},q_k^{j},q_{k+1}^{i},q_{k+1}^j)\delta q_{k}^{i}\label{eqq2}\\&+F_{ij,d}^{+}(q_{k}^{i},q_k^{j},q_{k+1}^{i},q_{k+1}^j)\delta q_{k+1}^{i}.\nonumber
  \end{align}
  
  
 It is well known that, for the single agent case (see \cite{mawest} Section $4.2.1$), by finding the critical points of the discrete action sum $\mathcal{A}_d$ for the discretization of Lagrange-d'Alembert principle, and with external forces described by smooth functions $F:TQ\to T^{*}Q$, one obtains the forced discrete Euler-Lagrange equations \begin{align*}0=&D_2L^d_i(q_{k-1},q_k)+F^{+}_{d}(q_{k-1},q_k)\\+&D_1L^d_i(q_k,q_{k+1})+F^{-}_{d}(q_k,q_{k+1})\end{align*} for $k=1,\ldots,N-1$; with variations $\delta q_k$ satisfying $\delta q_0=\delta q_N=0$. Here we are considering the approximations for the external forces $F_d^{\pm}:Q\times Q\to T^*Q,$
\begin{align*} 
\int_{t_k}^{t_{k+1}}F(q(t), \dot{q}(t))\delta{q}(t)\; dt\approx & 
F^-_{d}(q_k, q_{k+1})\,\delta q_k\\&+F^+_{d}(q_k, 
 q_{k+1})\,\delta q_{k+1}.
\end{align*}

The forced discrete Euler-Lagrange equations define the integration scheme $(q_{k-1}^i,q_k^i)\mapsto(q_k^i, q_{k+1}^i).$ 
  
Note that $TQ^s$ can be discretized as $(Q\times
Q)^s$ . For a constant time-step $h\in\mathbb{R}^{+}$, a path $q:[t_0, t_N]\to Q^s$ is replaced by a discrete path $q_d=\{q_k\}_{k=0}^{N}$ where $q_k=(q_k^1,\ldots,q_k^s)=q_d(t_k)=q_d(t_0+kh)$. Let $C_d(Q^s)=\{q_d:\{t_k\}_{k=0}^{N}\to Q^s\}$ be the space of discrete paths on $Q^s$ and define the discrete action sum $\mathcal{A}_{d}:C_d(Q^s)\to\mathbb{R}$ by \begin{align}\mathcal{A}_d(q_d)=\sum_{i=1}^{s}&\Big(\sum_{k=0}^{N-1}L_i^{d,F}(q_k^i,q_{k+1}^i)-\sum_{j\in\mathcal{N}_i}\left(F_{ij,d}^{-}(q_k^i,q_{k+1}^{i})\delta q_k^i\right.\label{actionsum}\\
\qquad\qquad+&\left.F_{ij,d}^{+}(q_{k}^{i},q_{k+1}^{i})\delta q_{k+1}^i\right)\Big)\nonumber\end{align} where, to define $\mathcal{A}_d$, we are using \eqref{eqq2} and the fact that 
\begin{align}
\int_{t_k}^{t_{k+1}}\mathbf{L}_F(q(t),\dot{q}(t))\,dt=&\int_{t_k}^{t_{k+1}}\sum_{i=1}^{s}\Big(L_i(q_i(t),\dot{q}_i(t))\\
&\qquad+\frac{1}{2}\sum_{j\in\mathcal{N}_i}V_{ij}(q_i(t),q_j(t))\Big)\,dt\nonumber\\
\simeq&\sum_{i=1}^{s}L_i^{d,F}(q_k^i,q_k^j,q_{k+1}^{i},q_{k+1}^j)\\=:&\mathbf{L}^d_F(q_k,q_{k+1})\label{dld}
\end{align}with $\mathbf{L}^{d}_F:(Q\times Q)^{s}\to\mathbb{R}$. 
\begin{prop}\label{prop2discrete}
Let $\mathbf{L}^d
_F:(Q\times Q)^s\to\mathbb{R}$ be the discrete Lagrangian \eqref{dld}. A discrete path $q_d=\{q_k\}_{k=0}^{N}\in C_d(Q^{s})$ extremizes the discrete action $\mathcal{A}_{d}$ if and only if it is a solution for the forced discrete Euler-Lagrange equations for $\mathbf{L}^d
_F$, 
\begin{align}\label{eqpropdiscrete}D_1L^d_i(q_k^i,q_{k+1}^i)=&-D_2L^d_i(q_{k-1}^i,q_k^i)-\sum_{j\in\mathcal{N}_i}\left(D_1V_{ij}^{d}(q_k^i,q_k^j)\right.\nonumber\\&\left.\qquad\qquad+F^{+}_{ij,d}(q_{k-1}^i,q_{k-1}^j,q_k^i,q_k^j)\right.\\&\left.\qquad\qquad+F^{-}_{ij,d}(q_k^i,q_k^j,q_{k+1}^i, q_{k+1}^j)\right)\nonumber\end{align} for $k=1,\ldots,N-1$; $i\in\mathcal{V}$ and for variations $\delta q_k=(\delta q_k^{1},\ldots,\delta q_k^{s})$ satisfying $\delta q_0=\delta q_N=0$.
\end{prop}


Note that $V_{ij}^d$ only depends on $(q_{k}^i, q_k^j)$, but may instead depend on
 $(q_{k+1}^i, q_{k+1}^j)$, if a different discretization is used.

Equations \eqref{eqpropdiscrete} define a discrete flow, $
\Upsilon_{\mathbf{L}^d_F}\colon (Q\times Q)^{s}\to (Q\times Q)^s$, by $\Upsilon_{\mathbf{L}^d_F}(q_{k-1}, q_k)=(q_k, q_{k+1})$ where $q_k=(q_k^1,\ldots,q_k^s)\in Q^s$.


 \section{Simulation results}
 
Consider $s$ kinematic agents evolving on $Q=\mathbb{R}^n$ endowed with the Euclidean Riemannian metric, with local coordinates $q_i^A$, $A=1,\ldots,n$, and each one with unit mass, that is, $L_i=\frac{1}{2}||\dot{q}_i||^2$. We choose (\ref{eq: Vij}) as potential functions, and the external forces are given by $F_{ij}(q_i,q_j,\dot{q}_i,\dot{q}_j)=l_{ij}(\dot{q}_i-\dot{q_j})$ with $l_{ij}$ the entries of the Laplacian matrix $\mathcal{L}$ associated with the undirected graph $\mathcal{G}$. Denoting by $\Gamma_{ij}=(||q_i^A-q_j^A||^2-d_{ij}^{2})$ and using Proposition \ref{Th1}, the dynamics for shape control with flocking behavior is given by the following set of second-order nonlinear equations
\begin{equation}\label{eq18}
\ddot{q}_i^A=-\sum_{j\in\mathcal{N}_i}\left(\Gamma_{ij}(q_i^A-q_j^A)+l_{ij}(\dot{q}_i^A-\dot{q}_j^A)\right).
\end{equation}

Conditions on $\mathcal{G}$ and the set of desired distances $d_{ij}$ for \eqref{eq18} to achieve formation and velocity consensus are established in \cite{fl1}.

For the construction of the variational integrator, the velocities are discretized by finite-difference, i.e., $\displaystyle{\dot{q}_i=\frac{q_{k+1}^i-q_k^i}{h}}$ for $t\in[t_k,t_{k+1}]$. The discrete Lagrangian $L^d:(\mathbb{R}^n\times\mathbb{R}^{n})^{s}\to\mathbb{R}$ is given by setting the trapezoidal discretization for the Lagrangian $\displaystyle{\mathbf{L}(q,\dot{q})=\sum_{i=1}^{s}L_i(\pi_i(q),\tau_i(\dot{q}))}$, that is, \begin{align*}L_{i}^{d}(q_k^{i},q_{k+1}^{i})=&\frac{h}{2}L_i\left(q_k^i,\frac{q_{k+1}^i-q_k^i}{h}\right)\\&+\frac{h}{2}L_i\left(q_{k+1}^i,\frac{q_{k+1}^i-q_k^i}{h}\right)\end{align*} where, $h>0$ is the fixed time step. The discrete potential functions $V_{ij}^d$ are given by $V_{ij}^d(q_{k}^i,q_{k}^j)=\frac{1}{4}(||q_k^i-q_k^j||^2-d_{ij}^{2})^{2}.$

 
The external forces $F_{ij}(q_i,q_j,\dot{q}_i,\dot{q}_j)=-l_{ij} (\dot{q}_i-\dot{q}_j)$ are discretized also by using the trapezoidal discretization,  \begin{align*}
F_{ij,d}^{+}(q_{k-1}^{i},q_{k-1}^{j},q_k^{i},q_k^{j})&=\frac{l_{ij}}{h}((q_k^j-q_{k-1}^j)-(q_k^i-q_{k-1}^i)),\\
F_{ij,d}^{-}(q_{k}^{i},q_{k}^{j},q_{k+1}^{i},q_{k+1}^{j})&=\frac{l_{ij}}{h}((q_{k+1}^j-q_{k}^j)-(q_{k+1}^i-q_{k}^i)).
\end{align*}

Noting that the matrix $(I+h\overline{\mathcal{L}})$, with $I$ the identity matrix of proper dimensions, is always non-singular (see for instance \cite{berman} Theorem $2.3$) and by denoting $$\Gamma_k=\bar{B}D_z(I_{|\mathcal{E}|}\otimes\mathbf{1}_{s\times 1})\left(\left(I_{|\mathcal{E}|}\otimes\mathbf{1}\right)D_zz-d^2\right),$$ with $\bar{B}=B\otimes I_s$, $z=[||q_{ij}||^2]_{(2|\mathcal{E}|\times 1)}$ the vector of all relative positions, $D_z=\hbox{diag}(z)$, $D_zz=z^2$ where $z^2$ stands for the square of each component in the vector $z$, $d$ is the vector of all desired distances, $d^2$ denotes the square of each component of the vector $d$, and $\mathbf{1}_s$ is the $s$-dimensional vector with all entries equals to $1$, by using Proposition \ref{prop2discrete}, the forced discrete Euler-Lagrange equations are given by the explicit difference equation \begin{align}\label{eqqdiscrete2}q_{k+1}=&2(I+h\overline{\mathcal{L}})^{-1}q_k-(I+h\overline{\mathcal{L}})^{-1}(I-h\overline{\mathcal{L}})q_{k-1}\\&-\frac{h^2}{2}(I+h\overline{\mathcal{L}})^{-1}\Gamma.\nonumber\end{align} 


Note that the previous equations are a set of $ns(N-1)$ equations for the $ns(N+1)$ unknowns $\{q_k^i\}_{k=0}^{N}$, with $i=1,\ldots,s$. Nevertheless  the boundary conditions on initial positions and velocities of the agents $q_0^i=q_i(0)$, $v_{q_0}^i=\dot{q}_i(0)$ contribute $2ns$ extra equations that convert \eqref{eqqdiscrete2} into a nonlinear root finding problem of the $ns(N-1)$ equations and the same number of unknowns.

Equations \eqref{eqqdiscrete2} define the integration scheme by means of the 
discrete flow $\Upsilon_d:(\mathbb{R}^n\times \mathbb{R}^n)^s\rightarrow (\mathbb{R}^n\times \mathbb{R}^n)^s$ by
$$\Upsilon_d(q_{k-1},q_{k})=(q_{k},q_{k+1}),\quad q_k=(q_k^1,\ldots,q_k^s).$$

 To understand the rate of energy dissipation along the motion, note that because the discrete energy function is the discretization of the Hamiltonian function associated with the distance-based shape control problem with flocking behavior, the discrete energy shall be studied from a Hamiltonian formalism. Therefore, we will show in numerical simulations that the integration scheme $\Upsilon_d$ applied to the discrete Hamiltonian, i.e., $E^d_i\circ\Upsilon_d$ dissipates a low rate of energy compared with the classical explicit Euler, and while the agents are moving in formation it decays exponentially to zero. The same occurs for the total energy (i.e., the sum of the energies for each agent).


The total energy of each agent $E_i:
\mathbb{R}^{n}\times \mathbb{R}^{n}\to \mathbb{R}$ is given by $$E_i(q_i,\dot{q}_i)=\frac{1}{2}||\dot{q}_i||^2+\frac{1}{2}\sum_{j\in\mathcal{N}_i}V_{ij}(q_i,q_j).$$ Using the trapezoidal rule for $E_i$, the discrete energy function for each agent $E_i^d:\mathbb{R}^n\times\mathbb{R}^n\to\mathbb{R}$ is given by \begin{align*}E_i^d(q_k^i,q_{k+1}^i)=&\frac{1}{2h}(q_{k+1}^i-q_k^i)^2\\
&+\frac{h}{4}\sum_{j\in\mathcal{N}_i}((q_k^{i}-q_k^{j})^2-d_{ij}^{2})^{2}.\end{align*}


Next, to show the comparison of the variational integrator with an explicit Euler method, we consider planar agents, i.e. $Q=\mathbb{R}^2$ with a triangular formation with three agents.

The set of neighbors is given by $\mathcal{N}_1=\{2,3\}$, $\mathcal{N}_2=\{1,3\}$, $\mathcal{N}_3=\{1,2\}$. We choose the triangle defined by $d_{12} = d_{23} = d_{13} = 10$ as the desired shape.

To start the algorithm we use the ``correct'' boundary conditions for the first two steps, that is, 
\begin{equation*}
q_0^i = q_i(0),q_1^i=hv_{q_0}^i+q_{0}^i=h\dot{q}_{i}(0)+q_{i}(0).
\end{equation*}


We arbitrarily choose the following initial positions $q_0 = \begin{bmatrix}5.03 & -6.56 & 2.02 & 2.22 & -2.33 & 12.28 \end{bmatrix}$, and we set the initial velocities to be $\dot{q}_0 = \begin{matrix} [2.80 & -2.90 & 0.19 & 2.07 & -0.67 & 1.67]\end{matrix}$.
\begin{figure}[h!]
\centering

\includegraphics[width=0.6\columnwidth]{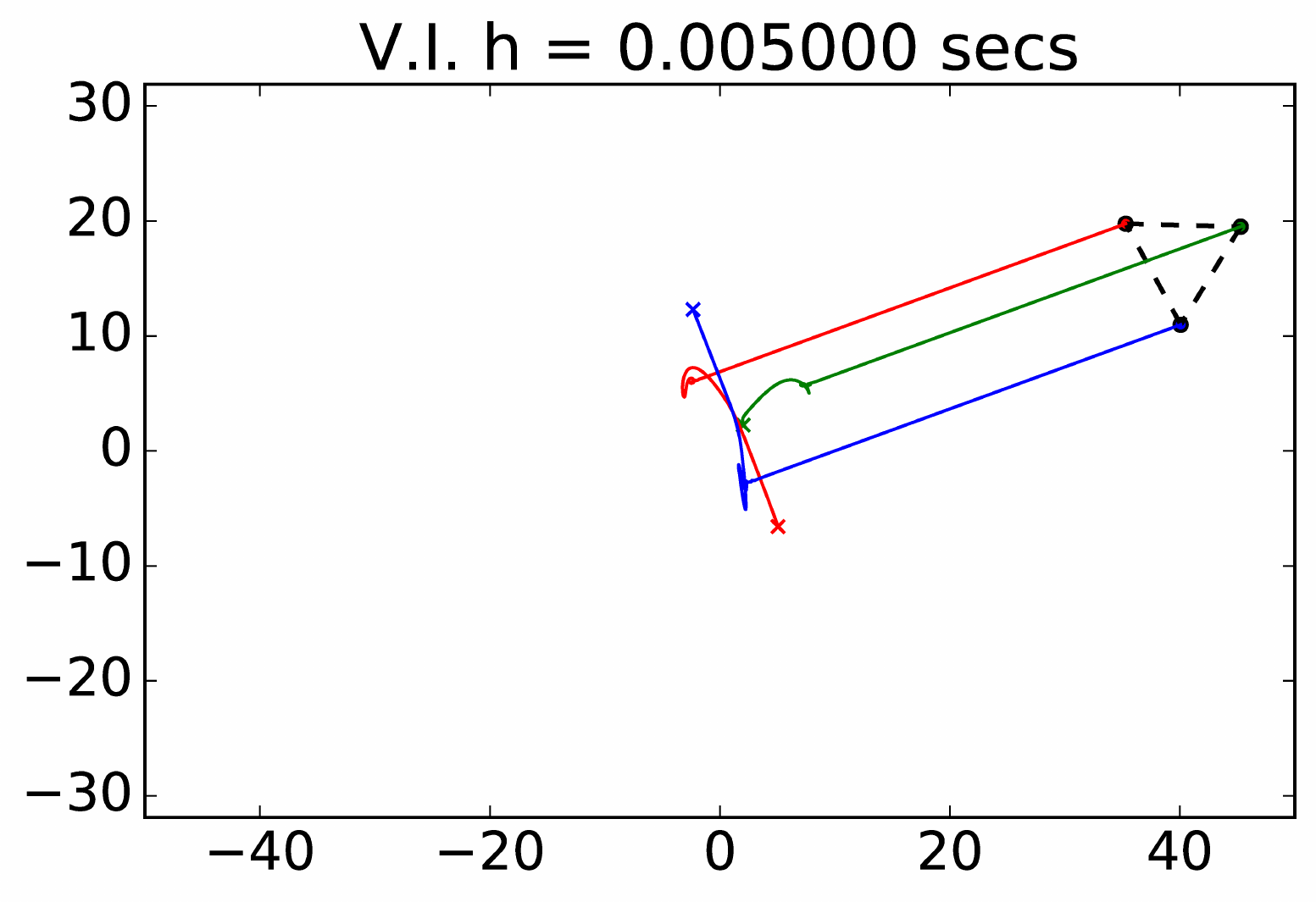}
\includegraphics[width=0.6\columnwidth]{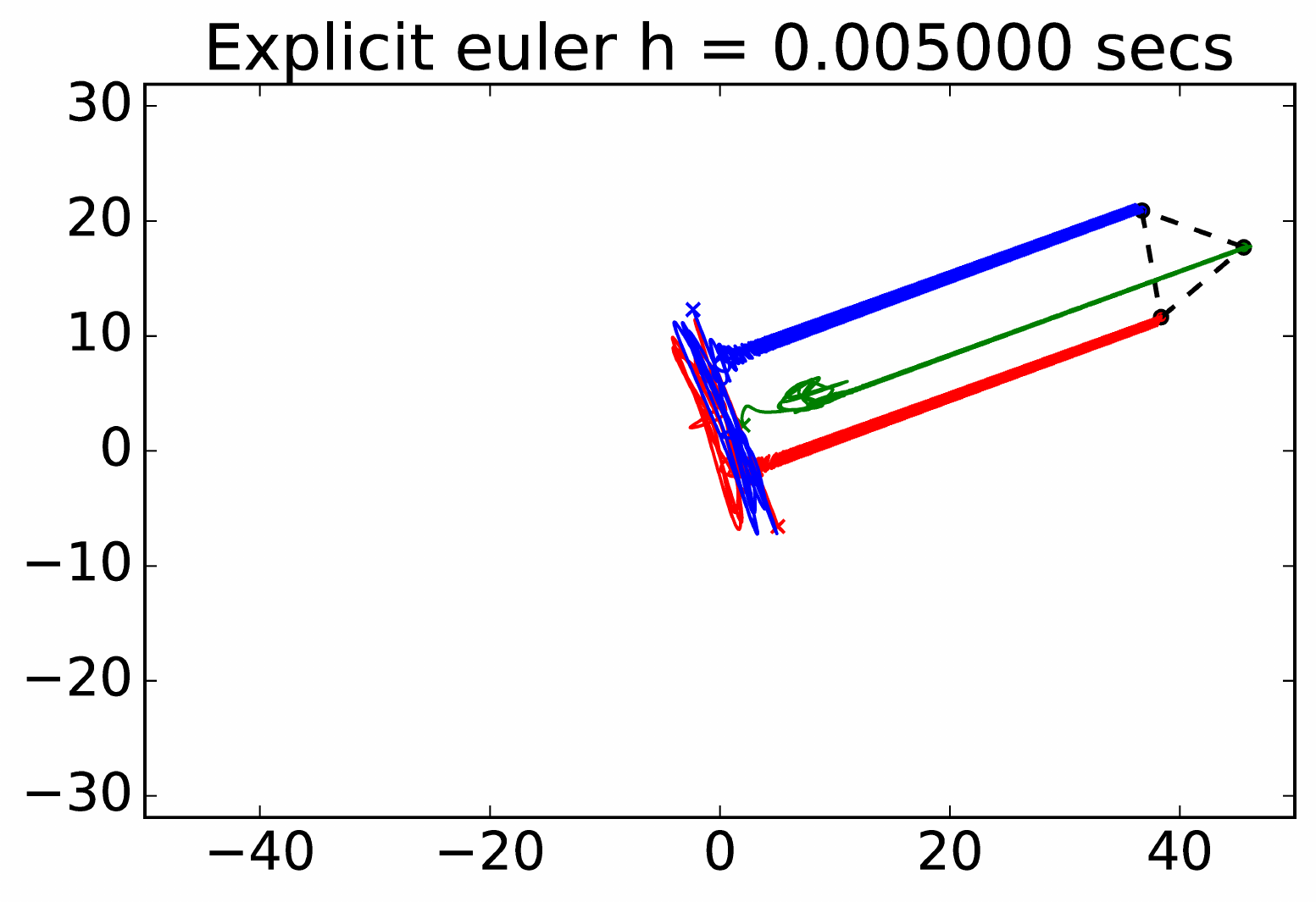}\\
\includegraphics[width=0.6\columnwidth]{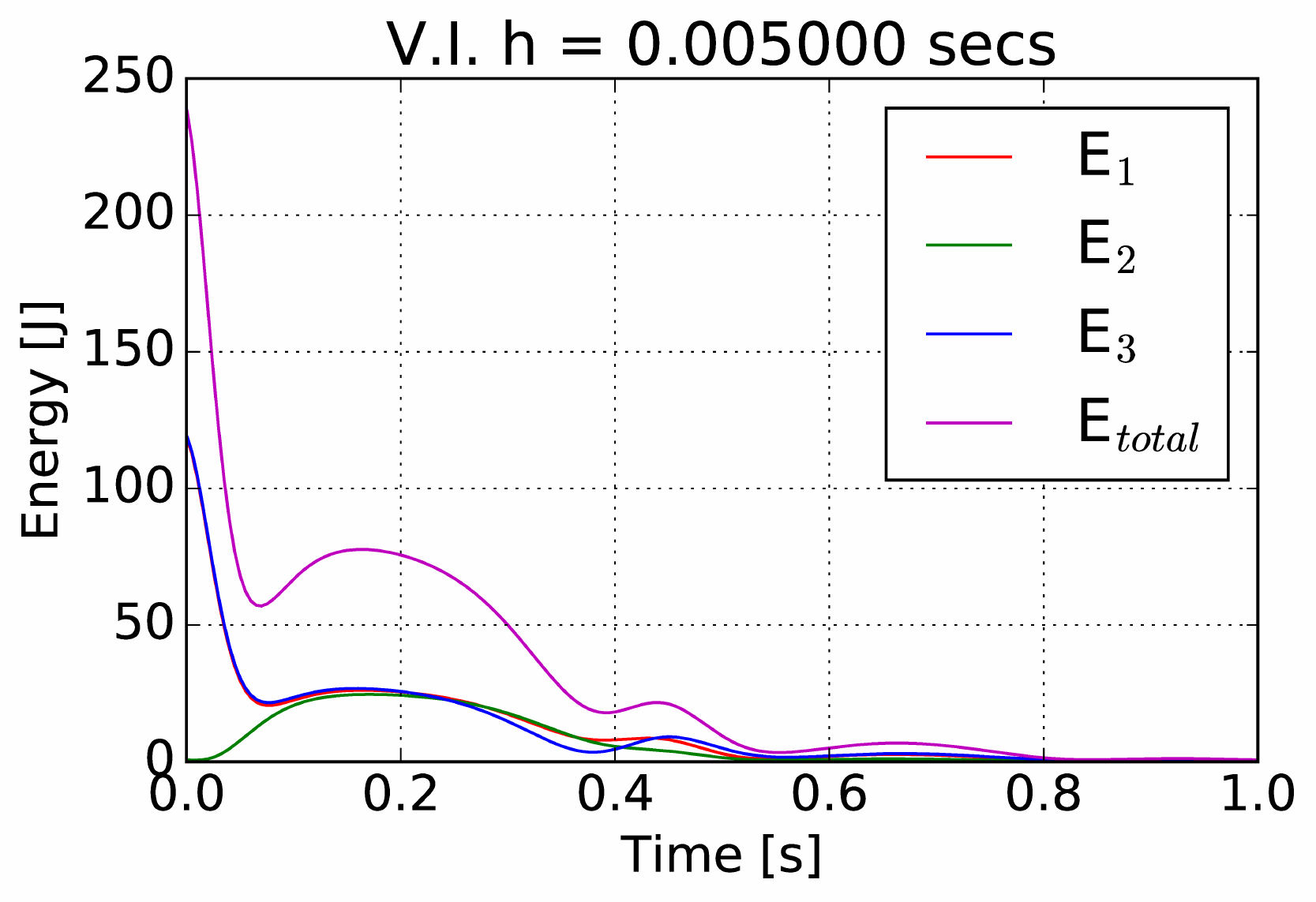}
\includegraphics[width=0.6\columnwidth]{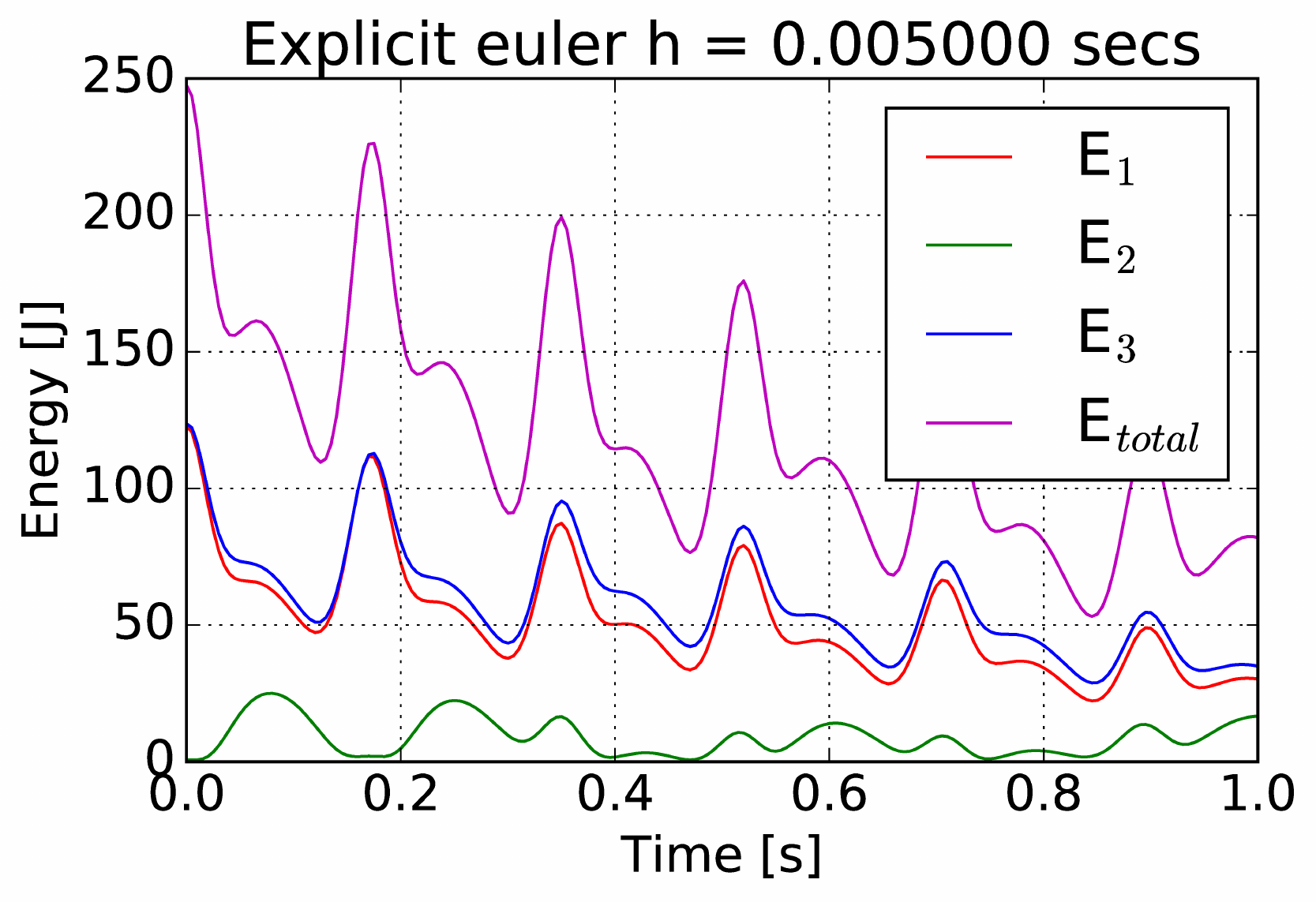}
\caption{ Agents' trajectories by employing the variational integrator (V.I.) and an explicit Euler method, with both having a fixed step size of $h=0.005$, and comparison between the discrete energy functions of the agents. The crosses denote the initial positions.}\label{fig1}
\end{figure}



 It can be verified that all the requirements on $\mathcal{G}$ and the set of $d_{ij}$ are satisfied so that \eqref{eq18} will achieve formation shape control and velocity consensus \cite{fl1}.

\begin{figure}[h!]
\centering
\includegraphics[width=0.6\columnwidth]{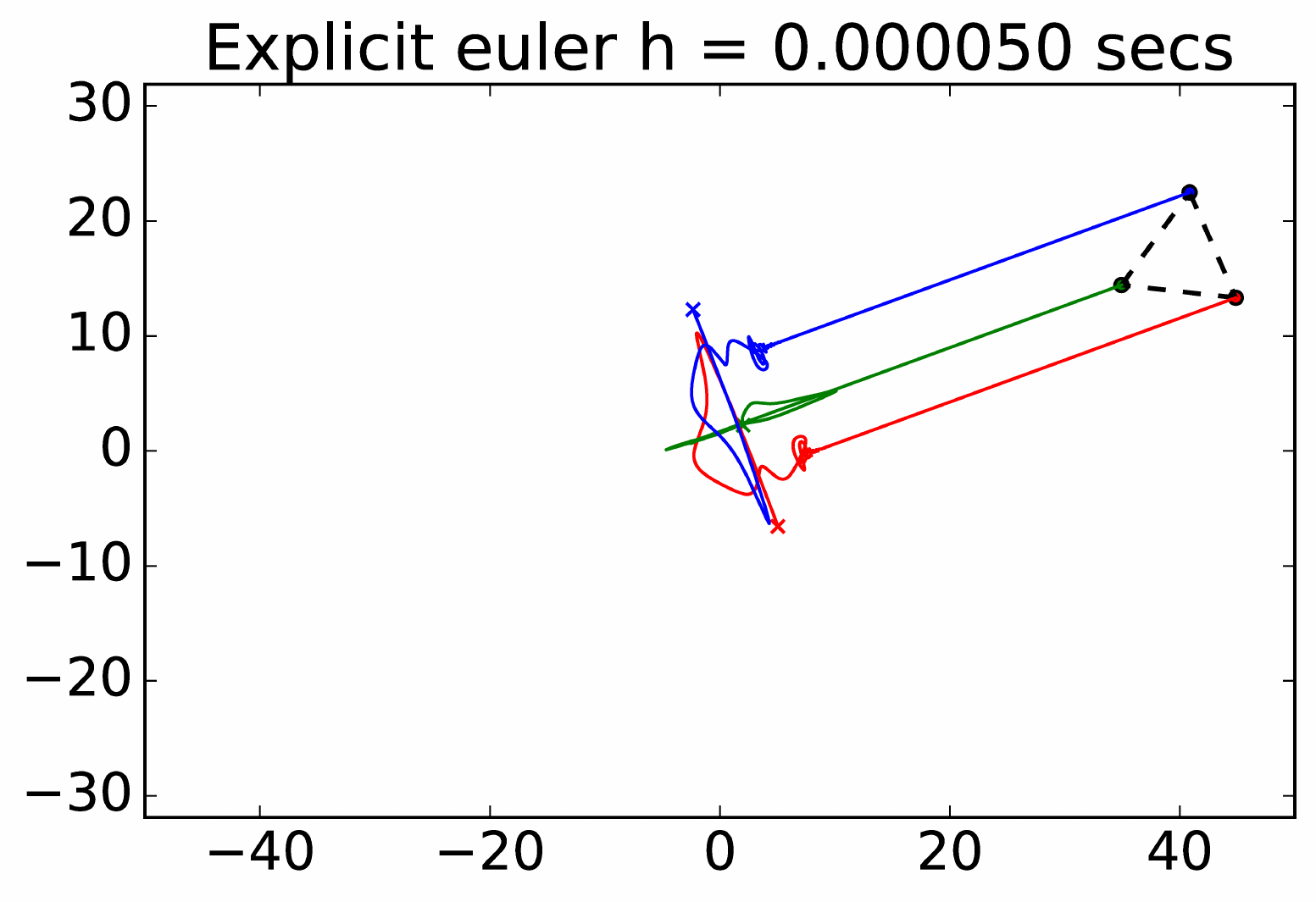}
\includegraphics[width=0.6\columnwidth]{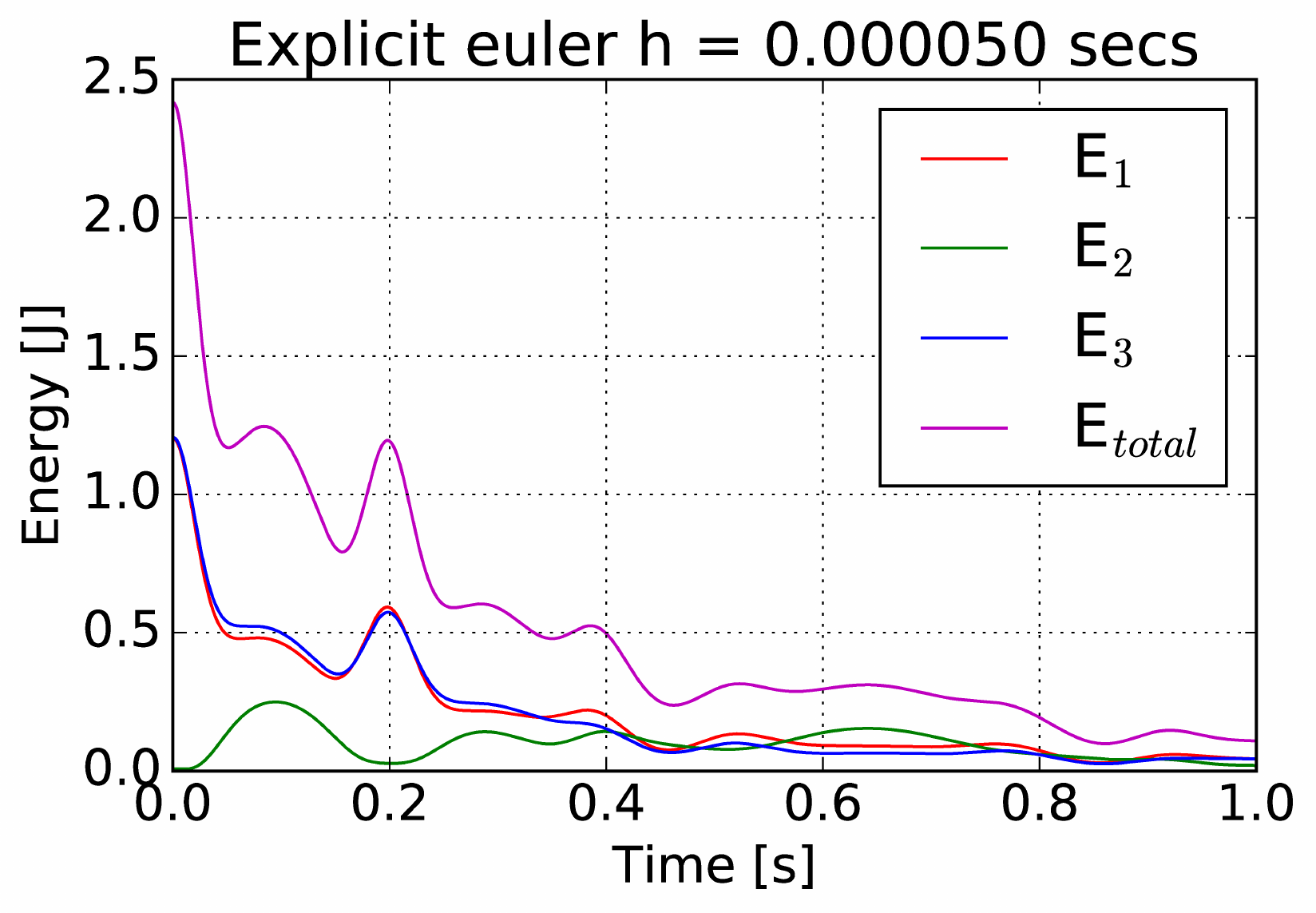}
\caption{Comparison between the discrete energy functions of the agents, the total energy (in black color), and the agents' trajectories by employing the variational integrator (V.I.) and Euler with both having a fixed step size of $h=0.00005$. The crosses denote the initial positions.}\label{fig3}
\end{figure}

Figure \ref{fig1} shows that the transitory and final formation shape are notably different between the proposed variational integrator and an explicit Euler-method. Specifically, the explicit Euler method generates unrealistic trajectories that differ greatly from the true trajectory of the continuous-time system \eqref{eq18}. Additionally, it can be seen in the lower figures that the energy dissipation in the explicit Euler approach is not accurate. Nevertheless with a lower $h$ (more steps of integrations) we achieve a similar result with the explicit Euler than with the variational integrator as we show in Figure \ref{fig3}. We have a consistent transitory with the explicit Euler method (and final desired shape) when we choose $h = 0.00005$ seconds or lower. We can therefore conclude that the simulation of \eqref{eq18} using the variational integrator \eqref{eqqdiscrete2} yields greater benefits compared to the explicit Euler method in terms of generating accurate trajectories, ensuring important physical properties such as the dissipation of energy has a good behavior, and lower computational cost.

In distance-based shape control, the desired shape is in general only locally stable.\footnote{For the triangular case it is almost globally asymptotically stable} The advantages in the performance of the variational integrator compared with classical integrator schemes is crucial, for instance, to develop an accurate and fast simulation to numerically determine regions of attraction to the desired final shape in swarms, as well as, to develop more computationally efficient estimation algorithms like Kalman filters that employ distance-based controllers as prediction models.

The methods and results here presented will help to numerically study and validate more complex formation control algorithms. In particular, when in practical applications we need to deal with the motion control of the formation and inconsistent measurements as it is shown in \cite{de2018taming}, or cases where a formation leader is specified, as in \cite{fl1}. Moreover, the proposed formation control is a distance-based one. As pointed out in \cite{mou}, mismatch in distance measurements may cause dramatic misbehaviors of multi-agent formation.  For future work we plan to study more complex systems, including non-point mass agents and analyze the design of geometric integrators that may be applied to study the effect of mismatches in distance measurements.

\bibliography{refs}

\end{document}